\documentclass[%
reprint,
amsmath,amssymb,
prl,
]{revtex4-1}

\usepackage{graphicx}% Include figure files
\usepackage{float}
\usepackage{dcolumn}% Align table columns on decimal point
\usepackage{bm}% bold math
\begin{document}

\preprint{APS/123-QED}

\title{High efficiency energy extraction from a relativistic electron beam \\ in a strongly tapered undulator}

\author{N. Sudar, P. Musumeci, J. Duris, I. Gadjev}
\affiliation{Particle Beam Physics Laboratory, Department of Physics and Astronomy University of California Los Angeles\\Los Angeles, California 90095, USA
}%

\author{M. Polyanskiy, I. Pogorelsky, M. Fedurin, C. Swinson, K. Kusche, M. Babzien }
\affiliation{
Accelerator Test Facility Brookhaven National Laboratory\\
Upton, New York 11973, USA% with \\
}%

\author{A. Gover}
\affiliation{Faculty of Engineering, Department of Physical electronics, Tel-Aviv University, Tel-Aviv 69978 Israel}

\date{\today}% It is always \toda y, today,
% but any date may be explicitly specified

\begin{abstract}
We present results of an experiment where, using a 200 GW CO2 laser seed, a 65 MeV electron beam was decelerated down to 35 MeV in a 54 cm long strongly tapered helical magnetic undulator, extracting over 30$\%$ of the initial electron beam energy to coherent radiation. These results demonstrate unparalleled electro-optical conversion efficiencies for a relativistic beam in an undulator field and represent an important step in the development of high peak and average power coherent radiation sources.
\end{abstract}

\maketitle

Greatly increasing the electro-optical conversion efficiency from particle beams to coherent radiation has the potential to enable a new class of high peak and average power sources capable of satisfying the increasing demands of cutting-edge scientific, defense and industrial applications. These range from powering laser-based accelerators, developing defense-class high energy lasers, and improving the throughput of next generation fabrication processes for the semiconductor industry\cite[]{LWFA,LAWS,EUV}.

The current workhorse to directly convert power from electron beams to electromagnetic radiation is the free-electron laser interaction where relativistic electron beams and electromagnetic waves exchange energy as they copropagate in an undulator magnetic field. This interaction is maximized when the electron energy, the undulator period and field amplitude satisfy the resonant condition, or equivalently the particles slip exactly one (or an integer number of) radiation wavelength every undulator period. In the classical FEL amplification scheme \cite[]{Madey,Pellegrini}, the amplification process saturates at a peak power given by $ P_{sat} \sim 1.6\rho P_{beam}$ where $\rho$ is the FEL pierce parameter (typically lower than 0.5 $\%$ for short wavelength radiation) and $P_{beam}$ is the beam power. Due to the absence of a gain medium or of a nearby metal or dielectric structure, the interaction is dissipation-free and the saturation occurs only due to the fact that the particles lose energy and fall out of the resonant interaction region.

Increasing the output power beyond the FEL saturation level can be achieved by tapering the undulator parameters --- that is modifying the undulator characteristics (field amplitude and/or period)-- to sustain the interaction even when the particles lose a large fraction of their energy. Undulator tapering as a means to increase FEL performances has been studied since the early days of FEL technology when the FEL was proposed as a path towards a very high average power source, and typically results in few percent efficiencies. The ELF experiment in the '80s demonstrated extraction efficiencies over 30 $\%$ but for GHz frequencies and only in a waveguide-mediated interaction \cite[]{Elf}. Recent development of the X-Ray FEL has rekindled interest in undulator tapering \cite[]{Fawley,Jiao,Emma} as increase in the peak power of the X-Ray FEL resulting from 5-10 $\%$ extraction efficiencies could unlock long-term goals in x-ray science such as single molecule imaging \cite[]{Chapman,Riege}.

An even stronger tapering of the undulator parameters to maintain the resonant condition over a very large (octave-spanning) beam energy variation has been studied in the context of Inverse Free Electron Laser (IFEL) accelerators \cite[]{Palmer,CPZ}. For example, the Rubicon IFEL at the Accelerator Test Facility at the Broohaven National Laboratory recently demonstrated resonant acceleration of particles from an initial energy of 52 MeV to a final energy of $\sim$ 95 MeV at a gradient of $\sim$100 MeV/m, \cite{DurisNatComms,ATFUM} using a 200 GW $CO_2$ laser pulse and a strongly tapered helical undulator.

In this letter, we discuss the results of an experiment operating such an accelerator in reverse, that is, where the high power $CO_2$ laser and the tapered helical undulator are used to obtain high gradient deceleration, halving the final beam energy, showing unprecedented efficiency in energy extraction from a highly relativistic electron beam. In the experiment, named Nocibur or inverse Rubicon, a permanent magnet based prebuncher was also used to bunch the electrons and load them at the decelerating phase of the interaction to maximize trapping efficiency. In summary, a fraction larger than 45$\%$ of the injected 65 MeV beam was decelerated to $\sim$35 MeV in the 54 cm long tapered helical undulator using a 200 GW 10.3 $\mu$m laser pulse showing for the first time the feasibility of reaching electro-optical energy conversion efficiencies as high as $30\%$ in short wavelength laser-electron interactions \cite[]{TESSA,Superradiance,patent}.

The reverse tapering of the undulator was determined using the resonant phase and energy concepts first introduced in Kroll, Morton and Rosenbluth \cite{KMR}. The electrons traveling in the undulator gain or lose energy depending on their phase in the ponderomotive potential defined by the laser and undulator parameters.  For the helical geometry employed in our experiments, the evolution of a particle energy is described by:
\begin{align}
\frac{d\gamma^2}{dz} &= -2 k K_l K \sin(\Psi)
\label{energy}
\end{align}
where $k$ and $k_w$ are the laser and undulator wavenumbers, $K_l=\frac{eE_0}{km_ec^2}$ and $K=\frac{eB_0}{k_w m_e c}$ are the laser and undulator vector potentials, and $\gamma$ and $\Psi$ represent the particle Lorentz factor and phase respectively. We define a resonant energy such that a particle at $\gamma_r$ will maintain a synchronous phase throughout the interaction, i.e.
\begin{align}
\frac{d\Psi}{dz} &= k_w - \frac{k(1+K^2)}{2\gamma^2} = 0 \rightarrow \gamma_r^2 = \frac{k(1+K^2)}{2k_w}
\label{phase}
\end{align}

To optimize the tapering (i.e. the variation of $k_w$ and $K$ and therefore of $\gamma_r$ along the undulator) we can derive a differential equation for the undulator parameters by equating the rate of change of the resonant energy defined by the resonance condition (i.e. the derivative of Eq. \ref{phase}) with the ponderomotive gradient expression (Eq. \ref{energy}) for a resonant particle at a constant non-zero resonant phase, $\Psi_r$ obtaining
\begin{align}
\frac{dK}{dz} &=\frac{(1+K^2)\frac{dk_w}{dz}}{2K k_w}-k_w K_l \sin{\Psi_r}
\label{K_tapering_prediction}
\end{align}

\begin{figure}[t]
\includegraphics[scale=0.4]{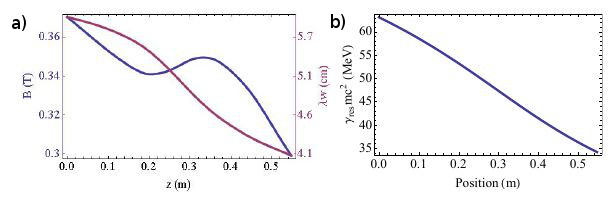}
\caption{(a) Undulator period ($\lambda_w=5.97 \rightarrow 4.04  cm$) and magnetic field amplitude tapering along undulator. (b) Resonant energy along the Nocibur undulator. }
\label{tapering}
\end{figure}

In the case of our experiment, the resonant phase $\Psi_r$ was set to $\pi/4$ as a compromise between the magnitude of the deceleration gradient and the extent of the stable region in longitudinal phase space where particles can be trapped and decelerated. Further, the variation of the period along the helical undulator which defines $\frac{dk_w}{dz}$ was pre-determined by the existing Rubicon undulator body and magnets (see Fig. \ref{tapering}a). The Nocibur experiment in fact re-utilized the Rubicon helical undulator made up of two period tapered Halbach undulators, oriented perpendicularly and shifted in phase by $\pi$/2 with period decreasing from 6 cm to 4 cm.  The undulator field amplitude was then adjusted to match the new field profile obtained as a solution of Eq. \ref{K_tapering_prediction} by varying the gap between the permanent magnets. Using the new undulator parameters, the resonant energy for 10.3 $\mu$m laser wavelength decreases from 65 MeV to 35 MeV along the interaction as shown in Fig. \ref{tapering}b.\\

\begin{figure}[b]
\includegraphics[scale=0.4]{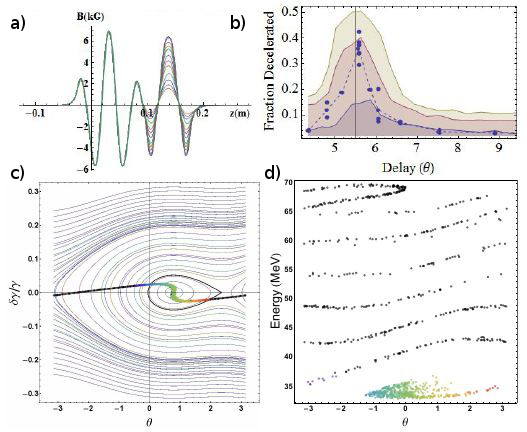}
\caption{(a) Hall probe measurements of the pre-buncher field varying chicane gap (b) Experimental data showing fraction captured varying pre-buncher chicane gap to control injection phase compared with GPT simulations with seed energy 0.55 J (yellow), 0.45 J (Red) and 0.35 J (Blue) (c) Longitudinal phase space for $\Psi_r = \pi/4$ IFEL ponderomotive potential with phase space curves for trapped and untrapped particles. The pre-bunched beam longitudinal phase space is also shown with the particles falling within the ponderomotive bucket separatrix color-coded.  (d) GPT simulation of the e-beam spectrum at Nocibur exit showing full deceleration for the pre-bunched particles. }
\label{lps_prebunching}
\end{figure}

In order to inject as many particles as possible in the stable region of the ponderomotive potential and therefore maximize the energy extraction efficiency we utilized a modulator-chicane compact pre-buncher. A single period planar Halbach-style permanent magnet based undulator with period 5~cm is used as an energy modulator. The chicane is formed by 4 dipole magnets of length 12.5~cm whose gap can be adjusted from a minimum of 13~mm to a maximum of 18~mm and interspaced by drifts of 12.5~cm. The prebuncher imparts a nearly 3 $\%$ peak-to-peak energy modulation on the beam. The variable gap allows us to control the dispersion of the chicane and tune the transport matrix element $R_{56}$ from 21 to 59 $\mu$m to obtain maximum compression.

Fine tuning of the prebuncher-chicane gap is used to control the relative injection phase between the laser and the electron microbunches at the entrance of the helical undulator Fig. \ref{lps_prebunching}a. Studying the fraction of the electron beam captured as a function of pre-buncher chicane gap, we observe a peak of maximum trapping where the electron beam is delayed by $\sim \frac{7\pi}{4} \lambda$, corresponding to slippage of the beam to the design resonant phase of $\pi$/4, Fig. \ref{lps_prebunching}b.\\

\begin{figure}
\includegraphics[scale=0.4]{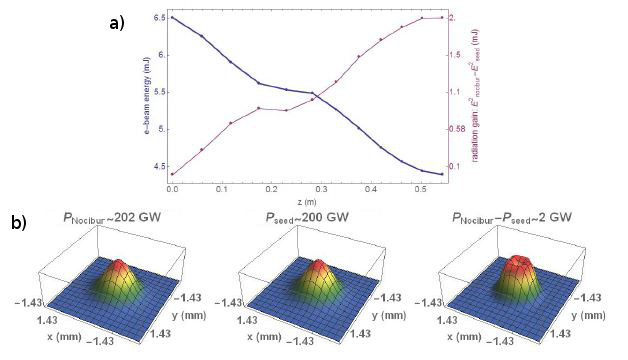}
\caption{(a) Genesis simulation data (200 GW seed, 100) A, 100 pC beam) showing energy gain of radiation and energy extracted from e-beam. (b) Transverse shape of Nocibur generated radiation at undulator exit.}
\label{genesis_sim}
\end{figure}

The full interaction was simulated in the General Particle Tracer (GPT) simulation code \cite{GPT} using field maps obtained from the 3D magnetostatic simulation code, Radia, agreeing well with hall probe measurements of the Nocibur undulator and pre-buncher.  Simulations of the radiation produced in the undulator were carried out using the 3D time-dependent FEL simulation code Genesis \cite[]{Genesis}.  Fig. \ref{lps_prebunching}c-d shows the longitudinal phase space distribution of the beam after the prebuncher as well as final electron beam spectrum and radiation growth and e-beam energy extraction from Genesis.  Without prebunching the fraction of particles trapped by the IFEL decelerator would have been less than 20 $\%$, reducing the extraction efficiency by a factor of 3 to $\sim10\%$. \\
\indent The output energy spectra from GPT and Genesis are in excellent agreement validating the assumption that minimal electromagnetic field evolution occurs along the interaction, Fig. \ref{genesis_sim}a. Genesis predicts an increase in radiation energy of 2 mJ, as expected given the energy lost by the electron beam and energy conservation. Detection of the generated radiation was hindered by the presence of the large signal from the drive laser pulse. By comparing the transverse profile of the seed pulse with the output pulse (Fig. \ref{genesis_sim}b) one notes that the new-born radiation comes out with a larger divergence angle as it should be expected since it is emitted by an electron beam focused in a much smaller spot size than the seed laser.

\begin{figure}[b]
\includegraphics[scale=0.45]{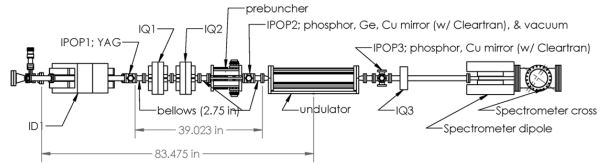}\\
\caption{\label{fig:wide} Nocibur experiment beamline layout.}
\label{layout}
\end{figure}

The amplitude of the radiation field plays a crucial role in maximizing the energy extraction efficiency and is important to highlight the difference between coherent undulator radiation emission and stimulated superradiant emission. This can be understood by considering the field generated by the passage of a bunched beam in an undulator magnet, $E_g$, emitted coherently with a high power seed field, $E_s$. The superposition of the two fields yields a total radiation pulse energy, $\varepsilon \propto (\overrightarrow{E_s}+\overrightarrow{E_g})^2 = {|E_s|}^2 +2 \Re [\overrightarrow{E_s}\cdot\overrightarrow{E_g^*}] +{|E_g|}^2$. The electromagnetic energy gained at the end of the undulator is then  proportional to $\Delta \varepsilon \propto 2 \eta_p E_s E_g \cos \phi +{|E_g|}^2$ where $\eta_p$ is the polarization matching factor (usually unity if the laser is properly circularly polarized) and $\phi$ is the phase of the bunching current relative to laser beam ($\cos \phi = \sin \psi_r = 1/\sqrt{2})$. The second term in this expression is the usual coherent undulator radiation. The first term represents the stimulated superradiant emission and for a large enough initial seed field can be dominating\cite{Gover:FEL}. For example in our case, if we calculate the coherent undulator emission from a perfectly microbunched 100 pC electron beam going through 11 periods of undulator we obtain 15 $\mu J$. Both simulations and as experimental energy spectra show instead mJ-level energy exchange between the particle and the radiation as a result of the stimulated interaction.

\begin{table}
\caption{\label{parameters} Parameters for the Nocibur experiment.}
\begin{ruledtabular}
\begin{tabular}{lcdr}
\textrm{Parameter}&
\textrm{Value}\\
\colrule
Initial electron beam energy & 65 MeV \\
Initial beam energy spread ($\frac{\Delta E}{E}$)& 0.0015\\
electron beam emittance ($\epsilon_{x,y}$)&2 mm-mrad\\
electron beam waist ($\sigma_{x,y}$)&100$\mu$m\\
electron beam current&100A\\
Laser wavelength & 10.3$\mu$m\\
Rayleigh range & 0.3 m \\
Laser waist & 990 $\mu$m \\
Laser waist position & $\frac{L_u}{2}$ = 0.225 m\\
Laser $M^2$ & 1.1\\
Laser Energy & 0.3 - 0.7 J\\
\end{tabular}
\end{ruledtabular}
\end{table}

Fig. \ref{layout} shows a detailed schematic of the beamline layout. The high power $CO_2$ laser seed is propagated through a 4.5 m focal length NaCl lens and coupled into the beamline through a NaCl window. Experimental electron beam and laser parameters are listed in Table \ref{parameters}. Picosecond scale timing between the laser and electron beam is achieved utilizing electron-beam controlled CO2 transmission in a semiconductor (Ge) sample \cite[]{Cesar:GeSwitch}. A delay stage is then finely adjusted to maximize the fraction of the electron beam decelerated.

%\begin{figure}[h]
%\includegraphics[scale=0.3]{prebfield2}
%\includegraphics[scale=0.175]{Nocibur_capture5}
%\caption{(Left) Hall probe measurements of pre-buncher magnetic field profile varying chicane gap. (Right) Experimental data %showing fraction captured varying pre-buncher chicane gap to control injection phase compared with GPT simulations with seed %energy 0.55 J (yellow), 0.45 J (Red) and 0.35 J (Blue).}
%\label{pb_capture}
%\end{figure}

\begin{figure}[t]
\includegraphics[scale=0.325]{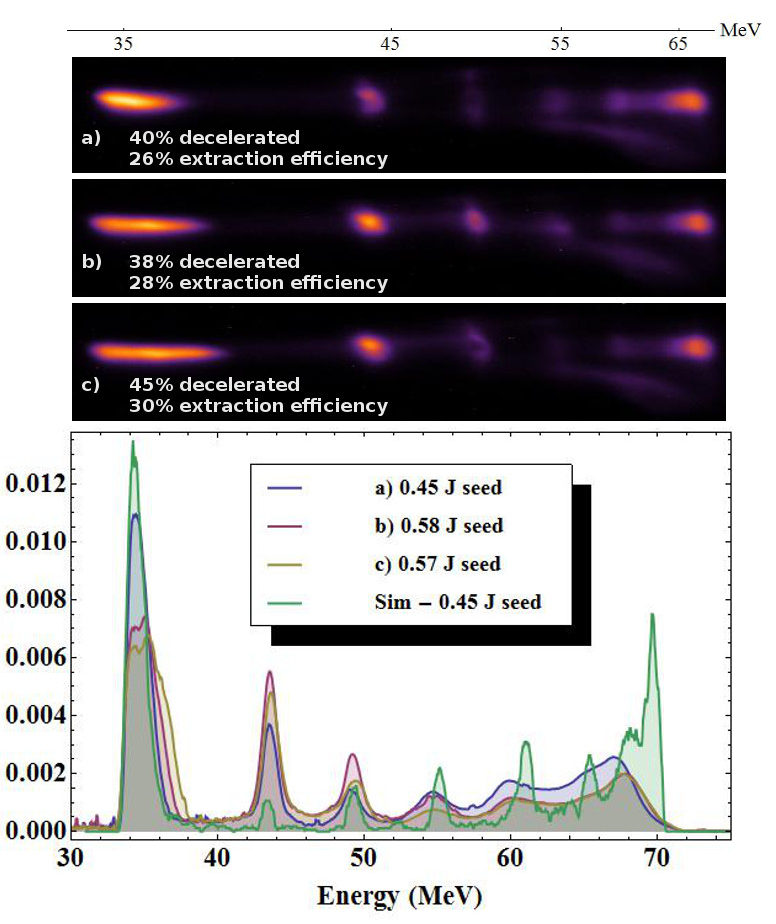}
\caption{Deceleration spectra from Nocibur spectrometer for 3 consecutive shots having slightly different input laser seed energy, compared with GPT simulation $\frac{1}{N}\frac{dN}{dE}$ vs. $E$ (Bottom).}
\label{spectra1}
\end{figure}

In Fig. \ref{spectra1} we show three representative energy spectrometer images and the relative projections to retrieve the energy distributions. The peak capture fraction was measured at $\sim$45\% for a 100 pC electron beam, injected at 65 MeV and decelerated down to 35 MeV, matching very well with the design simulations, Fig. \ref{spectra1}. The area under the shaded curves represents the total energy of the electron beam. The initial total energy in the beam was 65 MeV $\times$ 100 pC = 6.5 mJ. The total beam energy in the laser on shots can be obtained by integrating under the spectrum distribution curve to be 4.5 $\pm$ 0.4 mJ, from which we retrieve an energy extraction efficiency of $\sim 30\%$.

\begin{figure}[t!]
\includegraphics[scale=0.175]{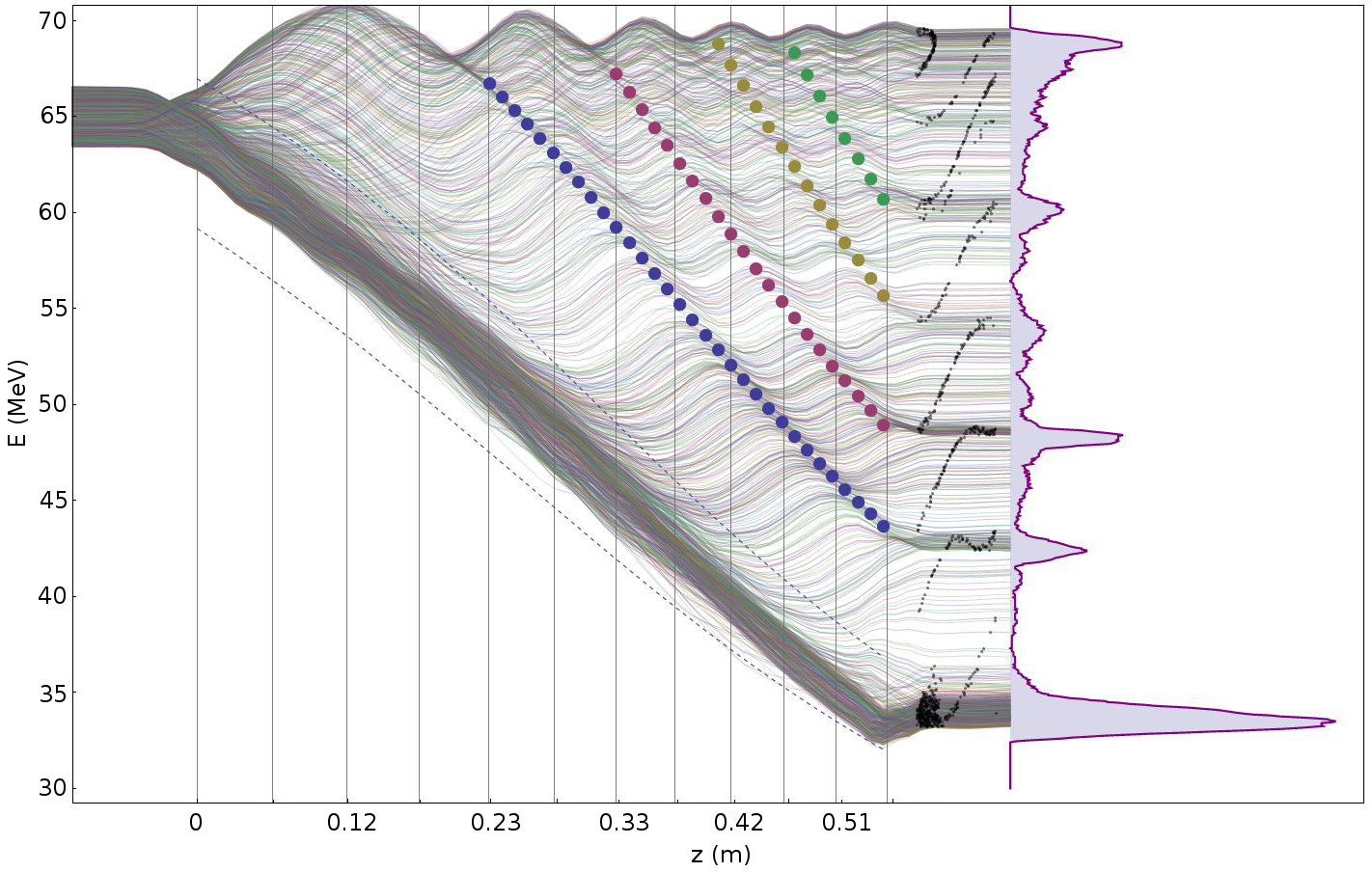}
\caption{Particle trajectories along the undulator from the GPT simulations. The ($\theta , \gamma$) longitudinal phase space at the undulator exit from Genesis simulation is displayed to show the remarkable agreement in all details of the energy spectrum. The ponderomotive potential bucket height is represented by dashed lines. The estimates for the positions of the detrapped energy peaks are also shown (points).}
\label{energypeaks}
\end{figure}

An interesting feature of both experimental and GPT simulation data is the discrete peaks in the energy spectrum of the detrapped particles. Analysis of the experimental and simulated spectra offer insights on the origin of this peaks as due to the characteristic non-resonant particle dynamics in the IFEL accelerator \cite{LCLSenergypeaks}. Much attention has been devoted in the literature to what happens to the particles trapped in the ponderomotive potential bucket, but an interesting effect is uncovered here for those electrons that follow the open trajectories in the phase space. Looking at Fig. \ref{lps_prebunching}c it is observed that for the particular resonant phase $\pi/4$ these trajectories 'bunch up' in energy at discrete levels. These energy levels can be calculated by finding the energy offsets for particles that have slipped ahead of the ponderomotive bucket by $2\pi n$.  Using the Hamiltonian defined in \cite[]{KMR} one can consider a detrapped particle, initially at $\delta\gamma=0$, and calculate the energy deviation after its phase slips by $2\pi$ yielding $\delta\gamma(z)\sim\sqrt{\gamma_r(z)\frac{d\gamma}{dz}\lambda_w(z)}$. The position along the undulator of the energy peaks at larger $\delta\gamma$ can be solved for numerically using the full Hamiltonian model. In Fig. \ref{energypeaks} we show representative trajectories for the particles along the undulator from the GPT simulation model in remarkable agreement with our estimates for the energy peaks using the Hamiltonian model. In principle, non-resonant IFEL interaction could find application in electron beam longitudinal phase space manipulation. For example by injecting in a tapered undulator a microbunched beam at a ponderomotive phase just outside the trapping bucket, it will be possible to take advantage of the phase space dynamics to stretch the beam and reduce its energy spread. \\

The results from the Nocibur experiment show 30 $\%$ electro-optical conversion efficiency from a relativistic electron beam opening the way for very high average and peak power radiation sources. Such high efficiency sets a new record for an interaction occurring between a free-space propagating short wavelength laser pulse and a relativistic electron beam comparing very favorably with early attempts to demonstrate high efficiency lasing in the far infrared regime \cite[]{Paladin}. This is mostly due to the developments in the generation of high brightness electron beams and increased seed laser quality and stability. The combination of using a prebunched beam, high intensity seed laser and strongly tapered helical undulator enable strong stimulated superradiant emission so that the radiated energy is orders of magnitude larger than coherent emission.  Finally, it should be noted that the experiment took advantage of the existing hardware and experimental setup from an ongoing IFEL accelerator experiment and so was not optimized for radiation generation resulting in emitted radiation significantly lower in intensity than the input seed. Nevertheless the experiment validates the approach and shows that large improvements in efficiency can be obtained when prebunched beams and strongly tapered undulators are used, thus opening the way to reach similar efficiencies with much larger amplification (high gain) at even shorter wavelengths where high intensity seed pulses are not available.

\end{document}